\documentclass[prl,twocolumn,showpacs,floatfix]
{revtex4}
\usepackage{graphics,epsfig,amsfonts,amssymb,amsmath,ulem,color}

\makeatletter

\renewenvironment{widetext@grid}{%
  \par\ignorespaces
  \setbox\widetext@top\vbox{%
   \vskip15\p@
   \hb@xt@\hsize{%
    \leaders\hrule\hfil
    \vrule\@height6\p@
   }%
   \vskip6\p@
  }%
  \setbox\widetext@bot\hb@xt@\hsize{%
    \vrule\@depth6\p@
    \leaders\hrule\hfil
  }%
  \onecolumngrid
  \let\set@footnotewidth\set@footnotewidth@ii
}{%
  \par
  \twocolumngrid\global\@ignoretrue
  \@endpetrue
}%

\makeatother

\def\e{\epsilon}

\def\s{\sigma}

\def\bk{{\bf k}}

\begin{document}
\title{Electronic correlations in Hund metals}
\author{L. Fanfarillo, E. Bascones}
\affiliation{Instituto de Ciencia de Materiales de Madrid, 
ICMM-CSIC, Cantoblanco, E-28049 Madrid (Spain).}
\email{leni@icmm.csic.es}
\date{\today}
\begin{abstract} 
To clarify the nature of correlations in Hund metals and its 
relationship with Mott physics we analyze the electronic correlations in 
multiorbital systems as a function of intraorbital interaction $U$, Hund's 
coupling $J_H$ and electronic filling $n$.  We show that the main process behind 
the enhancement of correlations in Hund metals is the suppression of the 
double-occupancy of a given orbital, as it also happens in the Mott-insulator at 
half-filling.  However, contrary to what happens in Mott correlated states the 
reduction of the quasiparticle weight $Z$ with $J_H$ can happen on spite of 
increasing charge fluctuations. Therefore, in Hund metals the quasiparticle 
weight and the mass enhancement are not good measurements of the charge 
localization. 
Using simple energetic arguments we explain why the spin 
polarization induced by Hund's coupling produces orbital decoupling. 
We also 
discuss how the behavior at moderate interactions, with correlations controlled 
by the atomic spin polarization, changes  at large $U$ and $J_H$  due to the 
proximity to a Mott insulating state.  
\end{abstract}
\maketitle

The Mott transition is one of the most dramatic manifestations of electronic 
correlations \cite{reviewmit, Fazekasbook}. In the single orbital Hubbard model at 
half-filling  the system becomes insulating at a critical interaction $U_c$ to 
avoid the cost of doubly occupying the orbital. Away from half-filling 
metallicity is recovered. Nevertheless atomic configurations involving double 
occupancy are avoided inducing strong correlations between the electrons. Charge 
fluctuations are suppressed and bad metallicity is observed. 

In multiorbital systems the Mott transition happens not only at half-filling but 
at all integer fillings \cite{rozenbergprb1997}. The crucial role of Hund's 
coupling $J_H$ on electronic correlations has been recognized only recently 
\cite{hanprb1998,wernerprl2008, shorikovarxiv2008, haulenjp2009, 
demediciprb2011,demediciprl2011,reviewhund,nevidomskyyprl2009, 
yinprb2012,akhanjeeprb2013,aronarxiv2014}. $J_H$ modifies $U_c$ in a doping 
dependent way \cite{hanprb1998,demediciprb2011} and promotes bad metallic 
behavior in a wide range of parameters \cite{haulenjp2009,demediciprl2011}. 

Within the context of iron superconductors, which accomodate 6 electrons in 5 
orbitals when undoped, the term Hund metal was coined  to name the correlated 
metallic state induced by Hund's coupling at moderate interaction 
$U$\cite{yinnatmat2011}. Originally Hund metals were described as strongly 
correlated but itinerant systems which are not in close proximity to a Mott 
insulating state and have physical properties distinctly different from doped 
Mott insulators \cite{reviewhund}. On the other hand, a number of authors 
\cite{ishidaprb2010, liebschprb2010, wernernatphys2011, nosotrasprb2012-2, 
lanataprb2013, demediciprl2014, nosotrasprb2014},  have described iron 
superconductors as doped Mott insulators due to the doping dependence of 
correlations observed: there is both experimental and theoretical evidence of an 
enhancement of correlations with hole-doping as the half-filling Mott insulator, 
with 5 electrons in 5 orbitals, is 
approached\cite{nosotrasreview2015,demediciprl2014,sudayamajpsj2012,hardyprl2013, 
terashimaprb2013,ishidaprb2010,liebschprb2010,wernernatphys2011, 
nosotrasprb2012-2,nosotrasprb2014,lanataprb2013}.

Orbital dependent correlations, named orbital differentiation, have been 
observed in some iron superconductors 
\cite{nosotrasreview2015,ishidaprb2010,yuprb2012, 
nosotrasprb2012-2,yiprl2013,terashimaprb2013,demediciprl2014} and are known to 
play an important role in ruthenates\cite{anisimovepjb2002}. It has been 
emphasized that Hund's coupling decouples the orbitals 
\cite{demediciprb2005,demediciprl2009,demediciprb2011, 
yuprl2013,nosotrasreview2015}, leads to orbital differentiation and even to an 
orbital selective Mott transition 
\cite{demediciprb2005,demediciprl2009,demediciprb2011}; however, the origin of 
such decoupling is not well understood. 

\begin{figure}
\leavevmode
\includegraphics[clip,width=0.485\textwidth]{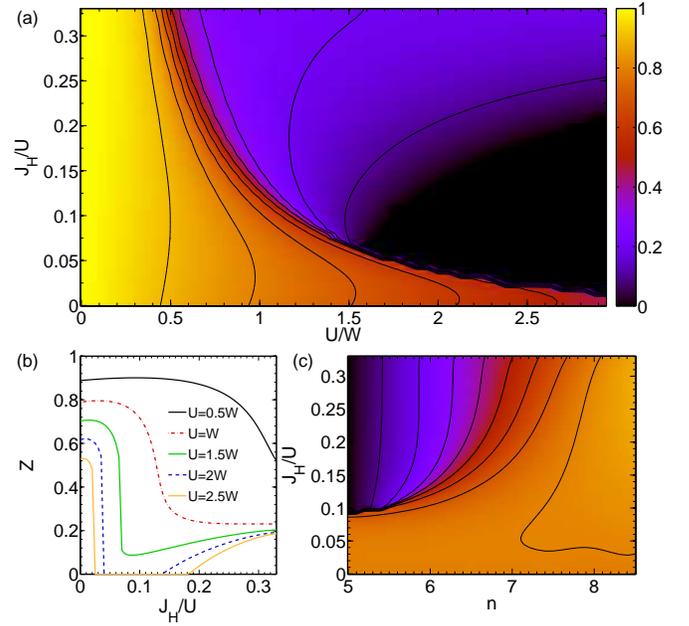}
\caption{(Color online) (a) Quasiparticle weight $Z$ vs intraorbital interaction 
$U$ and Hund's coupling $J_H$ for 6 electrons in 5 orbitals, the filling of 
undoped iron superconductors. $U$  and $J_H$ are in units of the bare bandwidth 
$W$ and $U$. A strongly correlated metallic region with small $Z$, in violet, 
appears in a wide range of parameters. The Mott insulating state is in black. 
The region in yellow-orange is metallic with moderate correlations. (b) $Z$ vs 
$J_H$ for the system in (a) and selected $U$. (c) $Z$ vs  electronic filling $n$ 
and $J_H$ with $U=W$ for a 5-orbital system. The strong suppression of $Z$ with 
$J_H$ seems connected to the $n=5$ half-filled Mott insulator.} 
\label{fig:n6}
\vspace{-.5cm}
\end{figure}
\begin{figure*}
\leavevmode
\includegraphics[clip,width=0.999\textwidth]{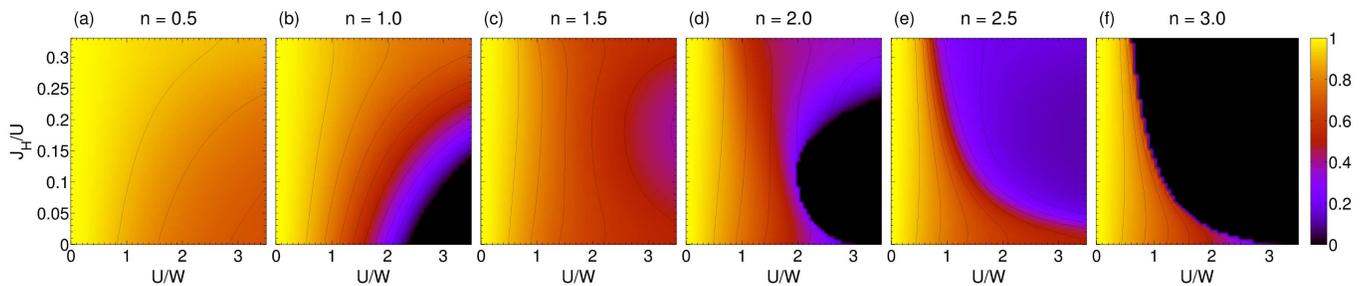}
\caption{ (Color online) Quasiparticle weight $Z$ vs intraorbital interaction 
$U$ and Hund's coupling $J_H$ for a 3- orbital system with electronic filling 
(a) $n=0.5$, (b) $n=1.0$, (c) $n=1.5$, (d) $n=2.0$ (e) $n=2.5$ and (f) $n=3.0$ 
half-filling. The Mott transition, in black, is found for the conmensurate 
values n=1,2,3 with different dependence on $J_H$. Extended metallic regions 
with strongly reduced $Z$ are only found for filling close to half-filling. $U$  
and $J_H$ are respectively given in units of the non-renormalized bandwidth $W$ 
and of $U$. The system shows particle hole symmetry, results are also valid for 
electronic filling $2N-n$.} 
\label{fig:Qp3orb}
\end{figure*}
It urges to clarify the nature of correlations in Hund metals and its 
relationship with Mott physics. In this paper we analyze the electronic 
correlations in multiorbital systems ($N=2,3...5$ orbitals) as a function of 
interactions and electronic filling $n$. We confirm that the doping dependent 
increase of correlations with $J_H$ at moderate interactions is directly 
connected to the Mott transition at half-filling.  However, contrary to what 
happens in correlated single-orbital systems the increase of correlations with 
$J_H$, as measured by the suppression of the quasiparticle weight $Z$, does not 
necessarily imply a suppression of charge fluctuations. We trace back this 
behavior to the opposite dependence of intra and interorbital charge 
fluctuations with Hund's coupling. With simple  
energetic arguments we explain the underlying phenomenology, including how the spin 
polarization drives the orbital decoupling.  Our study unveils differences 
between systems with 2 electrons and those with other commensurate partial 
fillings. We discuss a change of behavior at  large $J_H$ and $U$, related to 
the proximity of the Mott insulator. 

To address the generic features of Hund metals we consider degenerate 2D 
multiorbital systems with hopping $t$ restricted to the same orbital and to 
nearest neighbors and bandwidth $W=8t$.  For the interactions we start  from the 
Hubbard-Kanamori Hamiltonian \cite{castellaniprb1978,reviewhund}. 
\begin{eqnarray} 
\nonumber
H_{\text{int}}&=&U\sum_a n_{a\uparrow} n_{a\downarrow} 
+ (U'-J_H)\sum_{a< b,\sigma}n_{a\sigma}n_{b\sigma} \nonumber \\
&+&  U'\sum_{a\neq b}n_{a\uparrow}n_{b\downarrow} 
- J_H\sum_{a\neq b}c^\dagger_{a\uparrow}c_{a\downarrow}c^\dagger_{b\downarrow}c_{b\uparrow} \nonumber \\
&+&J'\sum_{a\neq b}c^\dagger_{a\uparrow}c^\dagger_{a\downarrow}c_{b\downarrow}c_{b\uparrow}
\label{eq:hamiltonian}
\end{eqnarray}
$a$ is the orbital index, $\uparrow$ and $\downarrow$ the spin, 
$n_{a\downarrow}$ and $n_{a\uparrow}$ the electron occupancy of a given orbital 
with spin $\downarrow$ or $\uparrow$. We treat the interactions using a Z$_2$ 
slave spin representation \cite{demediciprb2005,demediciprb2010}, and keep only 
density-density terms, see  Supplemental Material (SM). That is, pair hopping 
and spin-flip terms do not enter into the calculation 
\cite{demediciprl2014,nosotrasprb2014}. The model is particle-hole symmetric 
with respect to half-filling.  We take $U'=U-2J_H$, with $U'$ the interorbital 
interaction, as found in rotationally invariant systems 
\cite{castellaniprb1978}. Repulsive interactions    require $J_H/U\leq 0.33$.

The quasiparticle weight $Z$ provides a way to quantify the correlations between 
electrons.  $Z$ measures the overlap between the elementary excitations of the 
correlated and the non-interacting systems. It is equal to unity in 
non-interacting systems, decreases with increasing correlations and vanishes in 
Mott insulators. In Fermi liquid theory it equals the inverse of the mass enhancement. 
Fig.~\ref{fig:n6}(a) shows in color plot the quasiparticle weight 
$Z$ as a function of $U$ and $J_H$ for a five-orbital system with six electrons, 
the filling of undoped iron superconductors. Three regions can be distinguished: 
a metallic state with moderate correlations in yellow-orange color; an 
insulating Mott state at large $U$ in black, and a strongly correlated metallic 
state with reduced coherence in violet. The critical $U_c$ at which the Mott 
transition sets it depends non-monotonously on $J_H$ \cite{demediciprb2011}. At 
large values of $J_H$ the system remains metallic even for large $U$ 
\cite{demediciprl2011}. 

The correlated metallic state, in the following Hund metal, appears at finite 
$J_H$ in a wide range of parameters, including $U<W$. The way in which this 
region depends on the interactions reveals the crucial role played by $J_H$ on 
inducing the strong correlations which seem unrelated to the $n=6$ Mott 
insulating state. Similar phase diagrams are found in other cases, e.g. for 2 
electrons in 3-orbitals in Fig.~\ref{fig:Qp3orb}(d) and for 2 and 3 electrons in 4 
orbitals and 3 electrons in 5 orbital, in Fig.~S1 in SM.

Hund's coupling polarizes the spin locally.  The small $Z$ in a Hund 
metal is due to the small overlap between the non-interacting states and the 
spin polarized atomic states\cite{haulenjp2009,wernerprl2008,demedicichapter}. The suppression of $Z$ is thus concomitant with an 
enhancement of the spin fluctuations $C_S$, see Fig.~\ref{fig:fluctuations}(a). 
Here $C_S=<S^2>-<S>^2$ with $<S>=0$ and 
$S=\sum_{a=1,...,N}(n_{a\uparrow}-n_{a\downarrow})$. Arrows in 
Fig.~\ref{fig:fluctuations}(a) mark  $J_H^*(U)$ the interaction at which the 
system enters into the Hund metal defined empirically as the value of $J_H$ with 
the strongest suppression of $Z$, i.e. the most negative $dZ/dJ_H$ value, after 
which $Z$ stays finite, see Fig.~S2 in SM. Above $J_H^*$, $C_S$ reaches a value 
close to that of the Mott insulator at this filling, showing that in the Hund 
metal state each atom is highly spin polarized. 
\begin{figure*}
\leavevmode
\includegraphics[clip,width=\textwidth]{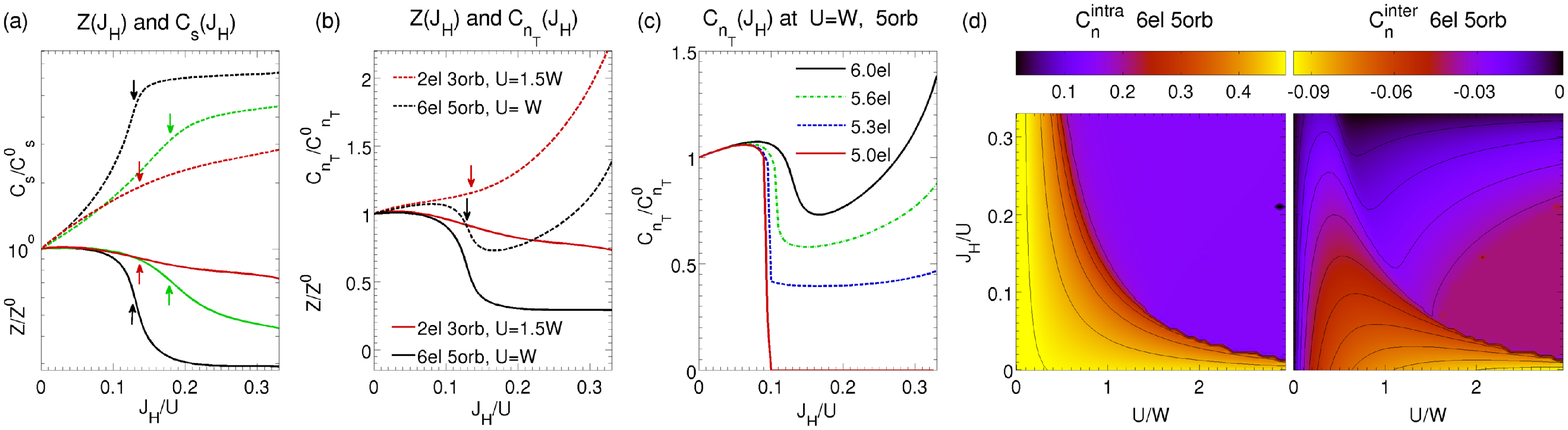}
\caption{ (Color online)(a) Enhancement of spin  fluctuations $C_S$ and 
suppression of $Z$ with $J_H$. $U=1.5W$ for 2 electrons in 3 orbitals (red), 
$U=W$ for 6 electrons in 5 orbitals (black) and for 3 electrons in 4 orbitals 
(green). $C_S$ and $Z$ are renormalized to their $J_H=0$ value at the given $U$. 
Arrows mark $J_H^*(U)$. The reduction of $Z$ is concomitant to the enhancement 
of $C_S$. (b) Charge fluctuations $C_{n_T}$ and quasiparticle weight $Z$ vs 
Hund's coupling $J_H$ renormalized to $C_{n_T}^0$ and $Z^0$, their value at 
$J_H=0$ and the corresponding $U$, see legend.  
The enhancement of $C_{n_T}$ while $Z$ is suppressed 
differs from the behavior of Mott correlated states. (c) $C_{n_T}/C_{n_T}^0$ vs 
$J_H$ for 5 orbitals with $U=W$ and different electronic fillings $n$. 
$Z$ decreases with $J_H$ for all values in this figure and vanishes in the Mott 
state at $n=5$ (not shown). (d) Intraorbital $C_n^{\text{intra}}$ and interorbital 
$C_n^{\text{inter}}$ charge fluctuations vs $U$ and  $J_H$ for 6 electrons in 5 
orbitals. With increasing $J_H$, both $C_n^{\text{intra}}$ and $C_n^{\text{inter}}$ 
decrease in absolute value. In the Hund metal $C^{\text{intra}}_n$ 
quickly saturates to its value in the Mott state while  $C^{\text{inter}}_n$ 
decreases towards zero with $J_H$. } 
\label{fig:fluctuations}
\end{figure*}

We now focus on the doping dependence of the correlations. Fig.~\ref{fig:n6}(c) 
shows $Z$ as a function of the electronic filling $n$ and $J_H$ for $U=W$, far 
from the $n=6$ Mott transition. The strength of correlations shows a clear 
asymmetry with electronic filling around $n=6$. No special feature is observed 
at $n=6$ for this value of $U$ what confirms that the $n=6$ Mott 
transition is not responsible for the strong suppression of $Z$. On the other 
hand the entrance to the strongly correlated Hund metal appears at smaller  
$J_H$ as $n$ approaches $n=5$. Connection with the Mott insulating state at 
half-filling is evident.

A clear doping dependence of correlations is also observed in 3-orbital systems, 
Fig.~\ref{fig:Qp3orb}. An extended region of parameters with small quasiparticle 
weight, in violet, is found only for fillings relatively close to half filling 
$n=3$. For smaller fillings $Z$ depends more weakly on $J_H$. The dependence of the 
Hund metal region on the interaction parameters for filling $n=2.5$ in 
Fig.~\ref{fig:Qp3orb}(e), closely follows the $n=3$ Mott insulating state, in 
black in Fig.~\ref{fig:Qp3orb}(f).

The hallmark of Mott physics is the suppression of charge fluctuations $C_{n_T}$ 
which vanish at the Mott transitions. Here $C_{n_T}=<n_T^2>-<n_T>^2=<(\delta 
n_T)^2>$ with  $n_T=\sum_{a=1,...,N}n_a$, $n_a=n_{a\uparrow}+n_{a\downarrow}$, 
$\delta n_T=n_T-<n_T>$  and  $<n_T>=n$.  In single orbital systems the charge 
fluctuations $C_{n_T}$ and $Z$ have a similar doping and interaction dependence. 
Consequently, very often, the suppression of $Z$, is assumed to imply 
localization.  

Fig.~\ref{fig:fluctuations}(b) shows the evolution of $C_{n_T}$ with $J_H$ and 
compares it with that of $Z$, both quantities being normalized to their 
$J_H=0$ value. Unexpectedly, $Z$ and $C_{n_T}$ depend differently
on $J_H$. For the system with 2 electrons in 3 orbitals $Z$ 
decreases and $C_{n_T}$ increases with $J_H$. That is, contrary to what happens 
in Mott systems, the suppression of $Z$ happens on spite of an increase of 
metallicity. In the 6 electrons in 5 orbitals case the strong reduction of $Z$  
comes along with a reduction of $C_{n_T}$. However at larger $J_H$, $Z$ 
continues decreasing,  while $C_{n_T}$ increases. The 
enhancement of $C_{n_T}$  with $J_H$ is reduced as half-filling ($n=5$) is 
approached, see Fig.~\ref{fig:fluctuations}(c).  The different dependence of $Z$ 
and $C_{n_T}$ on $J_H$ implies that in Hund metals the quasiparticle weight $Z$ 
and the mass enhancement are not good measures of the charge localization.

The increase of charge fluctuations with $J_H$ can be traced back to the 
suppression of interorbital correlations $C^{\text{inter}}_n$. Accounting for the 
equivalency of all the orbitals 
\begin{equation}
C_{n_T}=N \left (C^{\text{intra}}_n +(N-1)C^{\text{inter}}_n \right )
\label{eq:chargecorrelations}
\end{equation}
with $C^{\text{intra}}_n=<n_a^2>-<n_a>^2=<(\delta n_a)^2>$ the intraorbital 
fluctuations, $\delta n_a=n_a-<n_a>$ and $<n_a>=n/N$. $C^{\text{inter}}_n=<n_a 
n_b>-<n_a><n_b>=<\delta n_a \delta n_b>$ and $a \neq b$. $C^{\text{intra}}_n$, 
by definition positive or zero, is largest in the non-interacting limit.  
$C^{\text{inter}}_n$ is negative for repulsive interactions and it vanishes in 
the absence of interactions as the charge in different orbitals is not 
correlated. The entrance into the Hund metal has a very strong effect on 
$C^{\text{intra}}_n$ and  $C^{\text{inter}}_n$ being both strongly suppressed, 
see Fig.~\ref{fig:fluctuations}(d). Due to their different sign in 
Eq.~(\ref{eq:chargecorrelations}) this suppression has an opposite effect in 
$C_{n_T}$.  The increase of $C_{n_T}$ with $J_H$ is driven by the 
interorbital correlations which effect is enhanced by the 
degeneracy factor $(N-1)$ in Eq.~(\ref{eq:chargecorrelations}). On the other 
hand, the suppression of $C_{n_T}$ at $J_H^*$ in the 6 electrons in 5 orbitals 
case in Fig.~\ref{fig:fluctuations} is due to that of $C^{\text{intra}}_n$. 
Except at half-filling, $C^{\text{intra}}_n$ and $C^{\text{inter}}_n$ do not 
vanish in the Mott insulator but their contributions cancel each other leading 
to zero $C_{n_T}$, see Fig.~S3 in SM.
  
The phenomenology above can be understood by studying the energy of the hopping 
processes. Let's consider two $N$-orbital atoms with $n$ electrons ($n \leq N$) 
and assume that inside each atom the electron spins are parallel to satisfy 
Hund's rule. An electron which hops from one atom onto  the other one can end 
into (i) an empty orbital with spin parallel to that of the occupied orbitals 
with interaction energy cost $E^{\uparrow\uparrow}=U-3J_H$; (ii) an empty 
orbital with spin antiparallel to that of the occupied orbitals with $E^{inter 
\uparrow\downarrow}=U+(n-3)J_H$. (iii) an occupied orbital with 
$E^{intra\uparrow\downarrow}=U+(n-1)J_H$ \cite{notainteractions}. Particle-hole 
symmetry considerations apply for $n>N$.

At half-filling, $n=N$, processes (i) and (ii) are blocked by Pauli exclusion 
principle and process (iii) controls the critical $U_c(J_H)$ for the Mott 
transition which strongly decreases with $J_H$, see Fig.~\ref{fig:Qp3orb}(f). 
For other integer fillings and large  $J_H$ the Mott transition is controlled by 
process (i) and $U_c(J_H)$ increases with $J_H$ \cite{demediciprb2011},  but 
processes (ii) and (iii) are blocked at smaller interactions. 

In the metallic state the process (i) is allowed and promoted by $J_H$. We 
ascribe the entrance in the Hund metal at $J_H^*$ to avoiding process (iii). 
This process is suppressed by $J_H$ for  $n>1$ and it is directly connected to the Mott transition at half-filling. Its suppression strongly reduces the 
intraorbital double occupancy and $C^{\text{intra}}_n$ enhancing  the atomic 
spin polarization in the Hund metal. 

Process (ii) is suppressed by $J_H$ for $n>3$ and promoted for $n<3$ (as 
$U'=U-2J_H$). This introduces a qualitative difference between  systems with 2 electrons and those with larger integer fillings, which causes that  in 
2-electron systems the suppression of $Z$ and the enhancement of $C_S$ with 
$J_H$ are smoother and favors the enhancement of $C_{n_T}$. Nevertheless, even when process (ii) is allowed, the suppression of process (iii) indirectly reduces the occurrence of atomic configurations with anti-parallel spins in different orbitals (not shown).

The strong supression of $C_n^{intra}=<|\delta n_a|^2>$ at the crossover $J^*_H(U)$ reduces the interorbital charge correlations $<\delta n_a \delta n_b>$. The latter are further suppressed by the reduction of the effective interaction between the electrons in different orbitals, what produces orbital decoupling (measured by $\frac{<\delta n_a \delta n_b>}{<|\delta n_a|^2>}$): 
The interaction between electrons in different orbitals is $U'$ or $U'-J_H$ depending if they have parallel or antiparallel spins, see Eq.~(\ref{eq:hamiltonian}). At $J_H^*$ the occurrence of atomic configurations with parallel spin strongly increases while those involving opposite spin become less frequent, effectively reducing the interaction between electrons in different orbitals to $U'-J_H$ [\onlinecite{notadecoupling}]. If $J_H$ is further increased the decoupling is enhanced as the effective interaction $U'-J_H=U-3J_H$ decreases. At $J_H/U=0.33$, this interaction and $C^{\text{inter}}_n$ vanish.

As discussed above, at intermediate filling and interactions, around $J_H^*(U)$ 
the dependence of the quasiparticle weight and the fluctuations on the 
interactions is controlled by the establishment of the atomic polarization. On 
the other hand, at large $J_H$ and $U$, the behavior of the locally spin 
polarized system becomes dominated by the decrease with $J_H$ of the effective 
interaction between spin-parallel electrons and by the proximity to the Mott 
insulator, which happens at a larger critical interaction with increasing $J_H$. 
In particular, in the large $J_H$ and $U$ limit, both $Z$ and $C_{n_T}$ increase 
with $J_H$, see Figs.~\ref{fig:n6}(b), S2 and S3. Moreover while, with 
increasing $U$, $Z$ decreases monotonously, $C_S$ increase for intermediate $U$ 
but they start to decrease at large $U$ and $J_H$ see Fig.~S4. This behavior, 
driven by the interorbital spin fluctuations, is contrary to what happens in the 
single-orbital Hubbard model, for which $Z$ and $C_S$ show an opposite 
dependence on $U$ in all the range of parameters.  

In conclusion, we have clarified the nature of correlations in Hund metals and its differences with those in Mott systems. In Hund metals the enhancement of correlations originate in the suppression of atomic configurations which reduce the magnetic moment, specially intraorbital double occupancy,  while the hopping of electrons with spin parallel to the locally spin polarized atoms is allowed. The suppression of hopping processes involving intraorbital double occupancy links the correlations in Hund metals to the Mott transition at half-filling. However, contrary to what happens in Mott correlated systems,  the reduction of the quasiparticle weight $Z$ in Hund metals, can happen on spite of increasing charge fluctuations. Therefore in Hund metals the quasiparticle weight and the mass enhancement are not good measures of charge localization. 
The tendency towards orbital decoupling in the Hund metal is due to the reduction of the effective (and $J_H$-dependent) interaction between electrons in different orbitals produced by the predominance of atomic configurations involving parallel spins.
Finally, we note that at large $U$ and $J_H$ the dependence of the quasiparticle weight and the 
spin fluctuations on the interactions reveals a crossover 
to a region of parameters controlled by the proximity to the Mott insulator. 

The behavior discussed, together with other known properties of Hund metals \cite{demediciprb2011} 
as the enhanced width of the Hubbard bands or the screening of the atomic moments \cite{hansmannprl2010} 
is expected to play a prominent role in iron superconductors, ruthenates and many oxides. 
This is confirmed by the similarity between the behavior in Fig.1 and that found with realistic 
models of iron superconductors\cite{nosotrasreview2015,ishidaprb2010,yuprb2012}. Nevertheless the physics
of these materials will be strongly influenced by the inequivalency
of the orbitals\cite{nosotrasreview2015,ishidaprb2010,yuprb2012,nosotrasprb2012-2,
demediciprl2014,nosotrasprb2014,anisimovepjb2002} specific for
each material, and not included here. 

We thank L. de Medici for useful discussions and for providing us with the DMFT results used in the benchmark in the SM and to G. Kotliar, G. Giovanetti, Q. Si, R. Yu, A.J. Millis, R. 
Arita, M.J. Calder\'on, B. Valenzuela and the participants of the workshop 
"Magnetism, Bad Metals and Superconductivity: Iron Pnictides and Beyond" at KITP 
Santa Barbara for useful conversations. We acknowledge funding from Ministerio de 
Economia y Competitividad via FIS2011-29689 and FIS2014-53219-P from Fundaci\' 
on Ram\' on Areces, a fellowship from University of Rome La Sapienza and from 
the National Science Foundation under Grant No. NSF PHY11-25915. 
%\bibliography{hundmetal}
%\end{document}
\newpage
\mbox{}
\renewcommand{\thepage}{S\arabic{page}}  
\renewcommand{\thesection}{S\arabic{section}}   
\renewcommand{\thetable}{S\arabic{table}}   
\renewcommand{\thefigure}{S\arabic{figure}}
\renewcommand{\theequation}{S\arabic{equation}}

\setcounter{page}{1}
\setcounter{figure}{0}
\setcounter{table}{0}
\setcounter{equation}{0}
%\begin{document}
\title{Supplemental Material: Electronic correlations in Hund metal}
\author{L. Fanfarillo, E. Bascones}
\affiliation{Instituto de Ciencia de Materiales de Madrid, 
ICMM-CSIC, Cantoblanco, E-28049 Madrid (Spain).}
\maketitle

\begin{widetext}
\section{Supplemental Material: Supplementary figures}
\begin{figure*}[!h]
\leavevmode
\centering
\includegraphics[clip,width=0.90\textwidth]{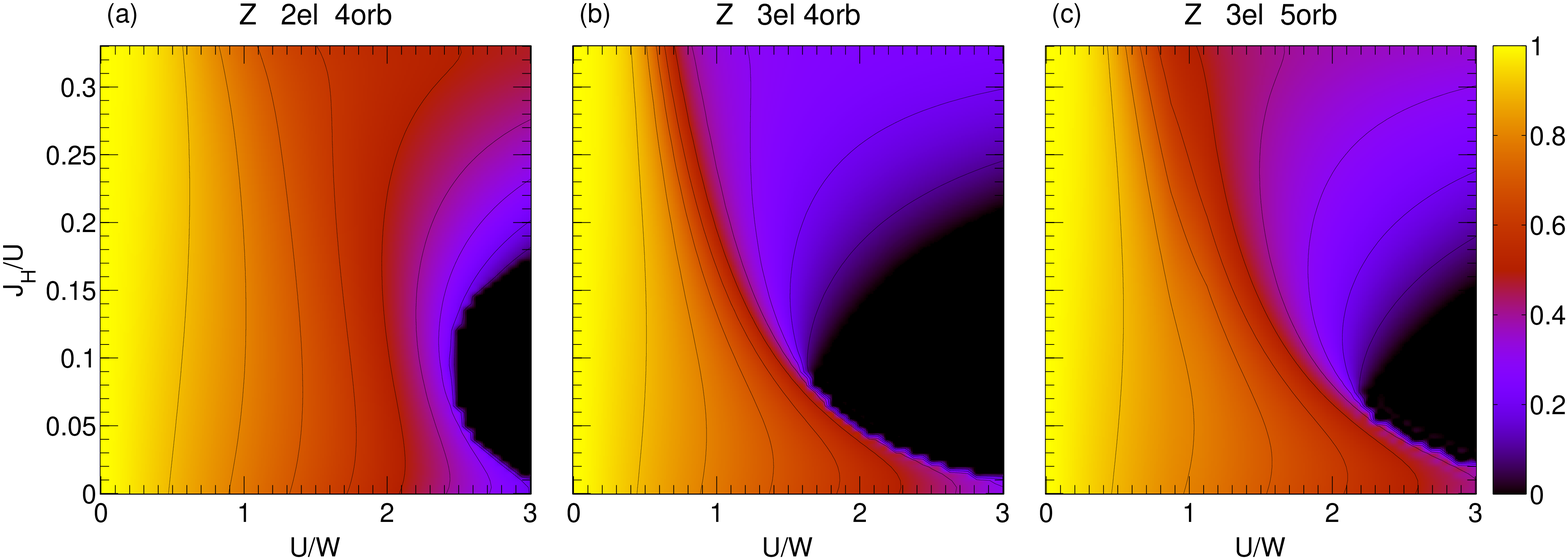}
\caption{ (Color online) Quasiparticle weight $Z$ as a function of intra-orbital 
interaction $U$ and Hund's coupling $J_H$ for (a) 4-orbital system with $n=2$ 
electrons , (b) 4-orbital system with $n=3$ electrons and  (c) 5-orbital system 
with $n=3$ electrons. $U$  and $J_H$ are respectively given in units of the 
non--renormalized bandwidth $W$ and of $U$. The system shows particle hole 
symmetry and the results are also valid for electronic filling $2N-n$. In (a) as 
in Fig.2(d) both with 2 electrons the Z contour lines for $U<2W$ are more 
vertical than in the other cases, i.e. weakly dependent on $J_H$. This is due to 
the different effect which for this filling has $J_H$ on the energy of the 
transport processes in which one electron hops into an empty orbital with 
opposite spin to that the atom,  see text.} 
\label{fig:supp1} 
\end{figure*}

\vspace{-0.3cm}
\begin{figure*}[h!]
\centering
\includegraphics[clip,width=0.90\textwidth]{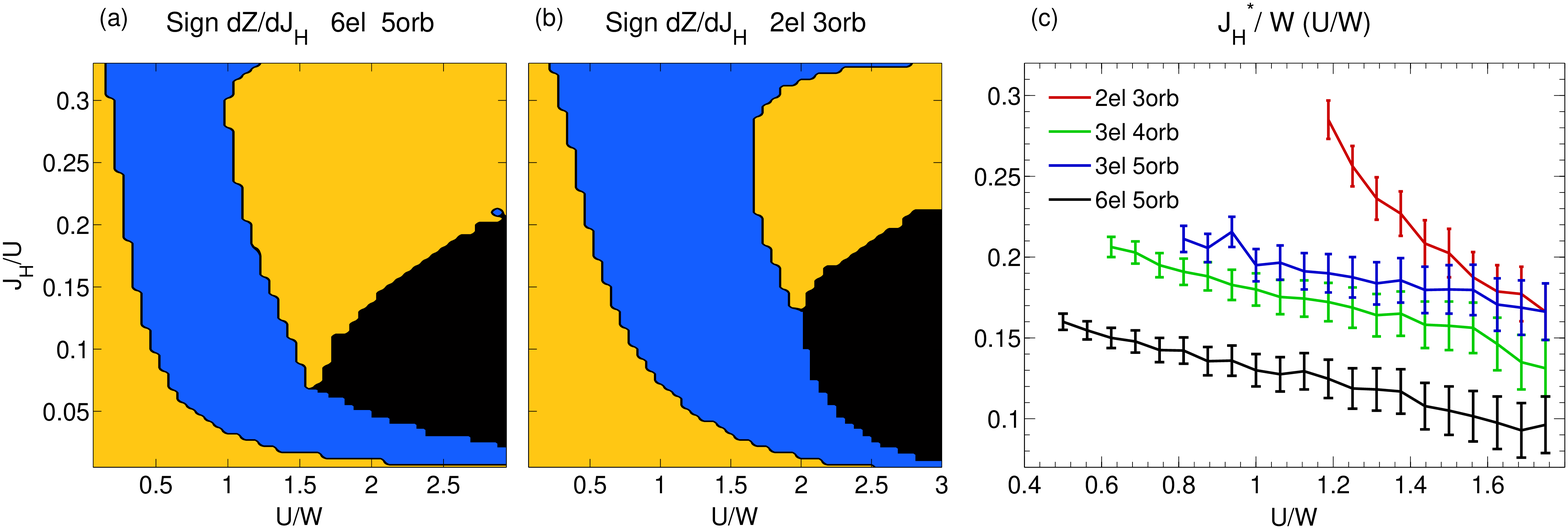}
\caption{ (Color online)(a) and (b) Color representation for the sign of the 
derivative of the quasiparticle weight with Hund's coupling $dZ/dJ_H$ for a 
5-orbital system with 6 or 4 electrons in (a) and a 3- orbital system with 2 or 
4 electrons in (b). Yellow is used for positive derivative, blue for negative 
derivative. In black, the Mott region with zero derivative. At intermediates 
values of $U$ and $J_H$, $Z$ is suppressed by $J_H$. At large $U$ and $J_H$, 
Hund's coupling promotes metallicity and $Z$ increases with $J_H$. The increase 
of $Z$ with $J_H$ found for small $U$ and $J_H$ is always very weak. (c) Hund's 
coupling $J_H^*$ at which the strongest suppression of quasiparticle weight $Z$ 
is found vs $U$. %From bottom to top: 6 electrons %in 5 orbitals (black),  3 
electrons in 4 (green) and 5 orbitals (blue) and 2 electrons in 3 orbitals 
(red). $J_H^*$ and $U$ are in units of the bandwidth $W$. Smaller $J_H^*$ values 
are found for smaller average orbital filling $x=n/N$, except when comparing 2 
electrons in 3 orbitals ($x=0.66$) and 3 electrons in 5 orbitals ($x=0.60$). }  
\label{fig:supp2}
\end{figure*}

\newpage

\begin{figure*}[h!]
\centering
\includegraphics[clip,trim= 0.1cm 0cm 0.cm 0cm, width=0.95\textwidth]{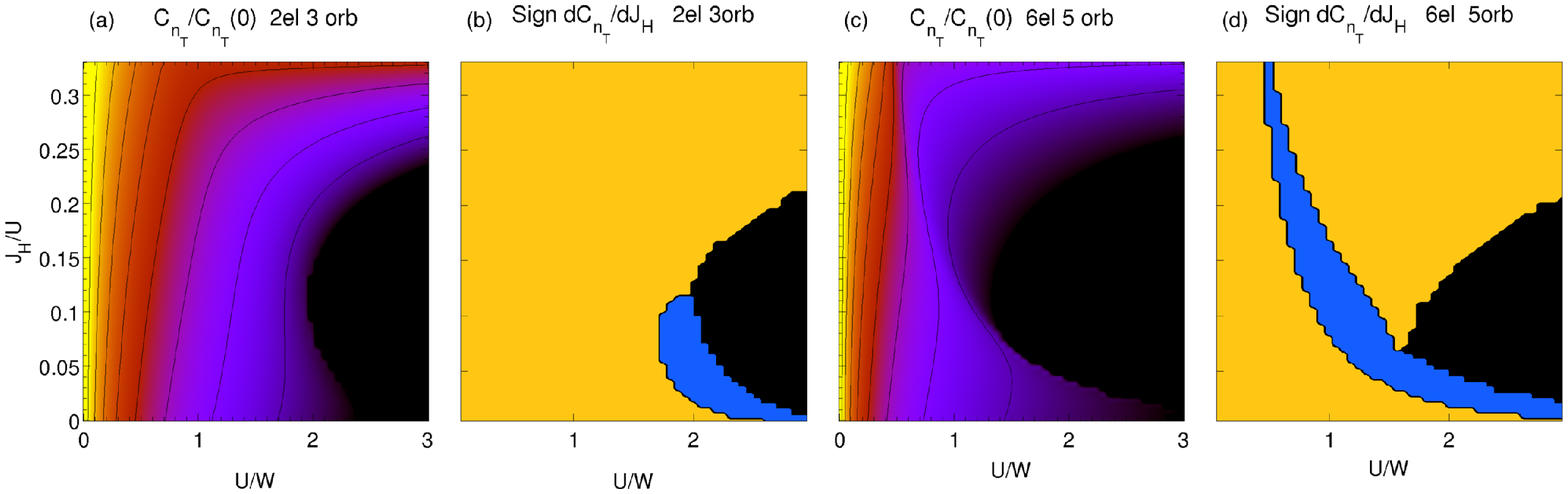} 
\caption{ (Color online) (a) and (c) Charge fluctuations $C_{n_T}$ normalized to 
its value in the non-interacting case ($U=0$ and $J_H=0$)  for 2-orbital system 
with 2 or 4 electrons in (a) and a 5-orbital system with 6 or 4 electrons in 
(c). This quantity varies between 0 and 1, facilitating the comparison with the 
dependence of $Z$ on interactions in Fig.~1 and 2 of the main text. (b) and (d) 
Color representation for the sign of the derivative of the charge correlations 
with Hund's coupling $dC_{n_T}/dJ_H$ for 2-orbital system with 2 or 4 electrons 
in (b) and a 5-orbital system with 6 or 4 electrons in (d), corresponding, 
respectively to the charge correlations in (a) and (c).  Yellow is used for 
positive derivative, blue for negative derivative. In black, the Mott region with 
zero derivative. In (d) there is a region of parameters with $dC_{n_T}/dJ_H<0$ 
which coincides with the region in  Fig.~ \ref{fig:supp2}(a) and Fig.~1 where 
the suppression of $Z$ with $J_H$ is strongest. In the case with 2 electrons in 
3 orbitals in (b) at moderate interactions charge correlations always increase 
with increasing $J_H$, even for values of the interaction with $dZ/dJ_H<0$. } 
\label{fig:supp3}
 \end{figure*}

\begin{figure*}[h!]
\centering
\includegraphics[clip,trim= 0.3cm 0cm 0.cm 0cm, width=0.94\textwidth]{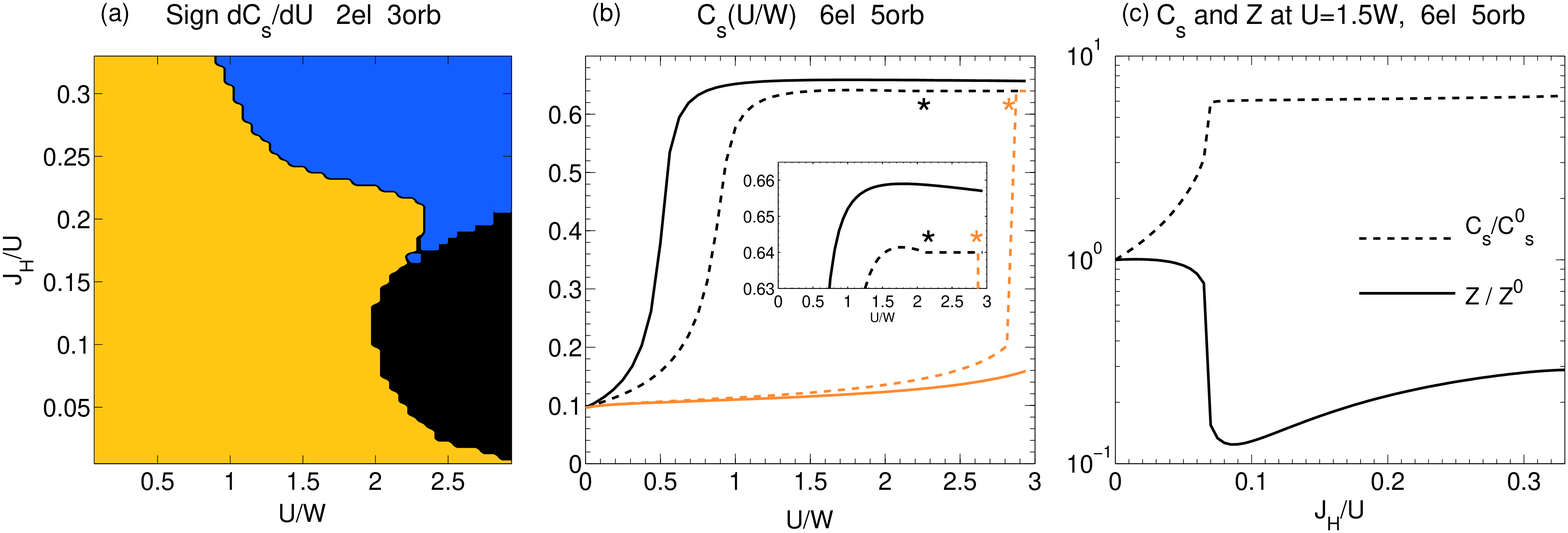}
\caption{ (Color online) (a) Derivative of the spin correlation $C_S$ with $U$ 
for a 3-orbital system with 2 electrons. At large $U$ and $J_H$, when the atoms 
are locally polarized, the spin fluctuations decrease with $U$ which promotes 
localization. The region where this effect is found is similar to that shown in 
Fig.~\ref{fig:supp2}(b) with positive $dZ/dJ_H$. (b) Spin fluctuations $C_S$ as 
a function of $U$ for 6 electrons in 5 orbitals and $J_H/U=0.30,0.15,0.015,0.01$ 
from top to bottom. For small $J_H/U$ $C_S$ increases with $U$. $C_S$ saturates 
in the Hund metal prior to the Mott transition, marked with stars.  Inset: Blow 
up of the large $J_H/U$ curves at large $U$ show non-monotonous $C_S$ with $U$ 
in this region of parameters.(c) Spin fluctuations $C_S$ and quasiparticle 
weight $Z$ normalized to their value at $J_H=0$ and the given $U=1.5W$ as a 
function of $J_H$. At $J_H^*$, $Z$ strongly decreases due to the enhanced spin 
polarization. However at larger values the behavior of these two quantities. 
$C_S$ saturates, while $Z$ starts increasing, due to the decrease of the 
effective interaction with $J_H$.} \label{fig:supp4}
\end{figure*}
 
 \begin{figure*}[h!]
\centering
\leavevmode
\includegraphics[clip,trim= 0.1cm 0cm 0.cm 0cm, width=0.90\textwidth]{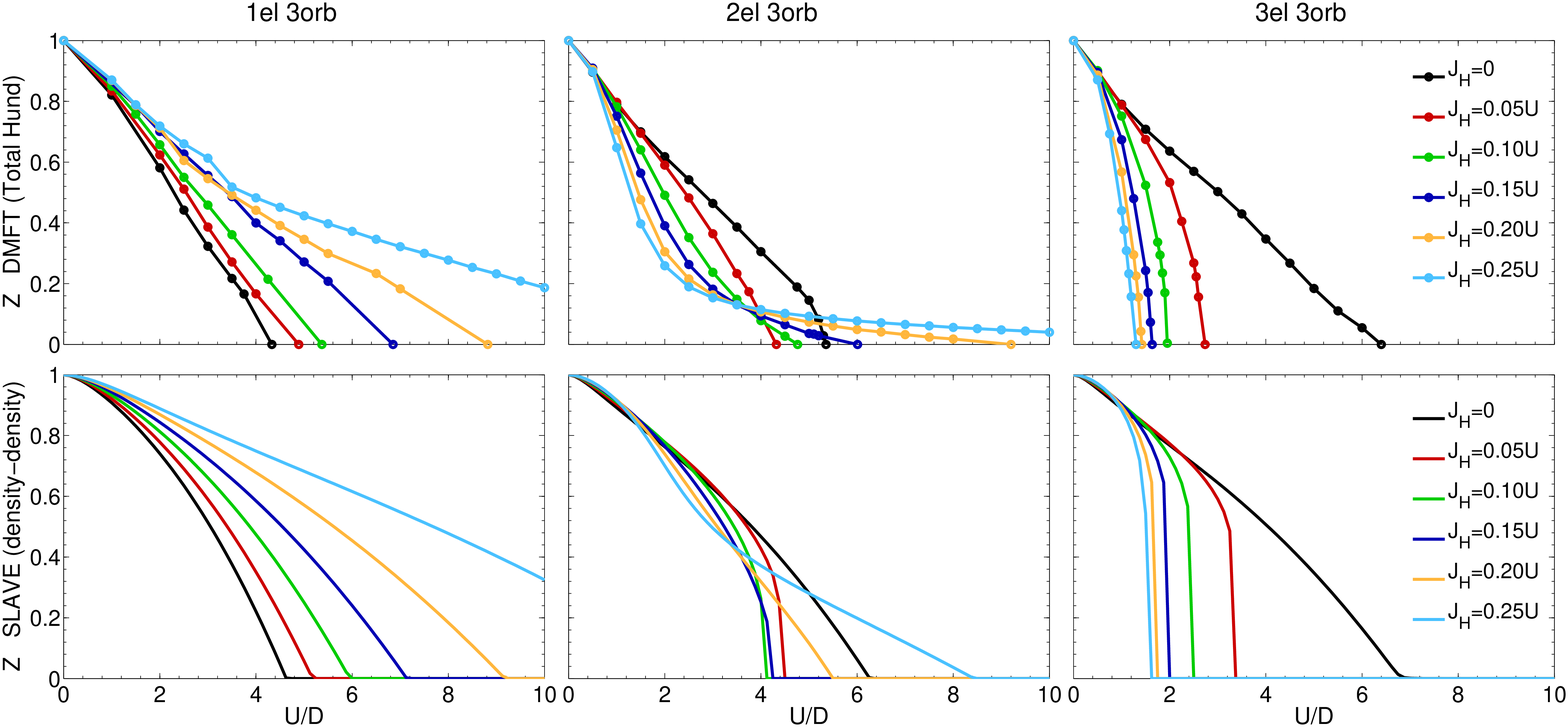}
\caption{
(Color online) Comparison between quasiparticle weight $Z$ of computed with DMFT 
and the Slave Spin method described. We show the  for a model with 3 equivalent 
orbitals with semicircular density of states with $n=1,2,3$ electrons per site. 
A semicircular density of states with half-bandwidth $D$ is used in both 
calculations. The DMFT data, taken from \cite{demediciprl2011},  were computed 
using the complete interacting Hamiltonian including the spin--flip and the 
pair--hopping terms. Only the density--density part (Ising approximation) is 
included in the slave-spin calculation.  
}
\label{fig:supp5}
\end{figure*} 
\end{widetext}

\section{Supplemental Material: The Ising $Z_2$ Slave-Spin approach}
Approaches which use slave particles have being widely used to deal with interacting fermions.
In particular, in multiorbital systems, slave--spin approaches, in $Z_2$ \cite{demediciprb2005,demediciprb2010} 
or $U(1)$ \cite{yuprb2012} versions, have proven to be very useful. 
In this work we have used the Z$_2$ slave-spin technique 
developed in \cite{demediciprb2005,demediciprb2010}
in its single-site approximation.  
 
In the $Z_2$ slave--spin approach the two possible occupancies of a spinless fermion on a given site, $n_c=0$ and $n_c=1$, 
are substituted by the two states of a 
pseudospin spin-1/2 variable, $S^z=-1/2$ and $S^z=+1/2$: 
$$
|0\rangle = |n_f=0,S^z=-1/2\rangle \quad\ \quad |1\rangle = |n_f=1,S^z=+1/2\rangle 
$$
$n_f$ is the occupation of an auxiliary fermion $f$, introduced to satisfy the anticonmutation relations. 
Two unphysical states 
$|n_f=0,S^z=+1/2\rangle, \, |n_f=1,S^z=-1/2\rangle$ are generated in the procedure. To eliminate them, the local constraint 
$
n^f = S^z + \frac{1}{2}
$
has to be imposed. 

In a multiorbital system each of the orbital and spin species have to be treated in this manner. That is, 
a set of $2N$ pseudospin-1/2 variables $S^z_{i a \sigma}$ 
and auxiliary fermions $f_{i a \sigma}$ are introduced at each site $i$ Here $a=1,N$ and $\sigma$ are the orbital and spin
indices. On each site these variables have to satisfy the local constraint:
\begin{equation}
n^f_{i a\s}=S^z_{i a \s}+\frac{1}{2},
\label{eq:constraint}
\end{equation}
what can be done with time-dependent Lagrange multipliers fields $\lambda_{i a\sigma}(\tau)$. 

Following the prescription in\cite{demediciprb2010} the physical fermions $c_{i a \sigma}$ are represented by
$$ c_{i a \sigma} = f_{i a \sigma } O_{i a \sigma}\, , \quad \quad  
c^\dagger_{i a \sigma} = f^\dagger_{i a \sigma }O^\dagger_{i a \sigma}. $$ 
Here $O_{i a \sigma}$  is a pseudospin-1/2 operator defined as
\[ O_{i a \s} = \left( \begin{array}{cc}
0 & \gamma_{i a \s}  \\
1 & 0  \end{array} \right)\] 
with $\gamma_{i a \s}$ a complex number\cite{demediciprb2010}, see below. 

To solve the interacting problem several approximations are introduced 
\cite{demediciprb2005,demediciprb2010}:  (i) Only the density-density terms of 
the Hubbard-Kanamori Hamiltonian, Eq.~(1) in the main text, are included 
\cite{notaspinflip}, (ii)  The constraint is treated on average, i.e. using a 
static Lagrange multiplier $\lambda_{a \sigma}$ and the Hamiltonians of the 
pseudospin slave variables and the auxiliary fermions are decoupled (iii) 
The problem is solved in a single-site mean field approximation, which render 
all variables site independent.  After these approximations the total 
Hamiltonian can be written as the sum of two effective Hamiltonians, for the 
auxiliary fermions and the pseudospins, $H_{f}$ and $H_{\text{PS}}$,  to be 
solved self-consistently at mean-field level. 

The fermionic Hamiltonian for a generic multiorbital system without orbital hybridization is: 
\begin{equation}
H_{f} = \sum_{a \, \sigma} \sum_{\bk}   \left( Z_{a \s } \, \varepsilon_{a \s} + \e_a - \mu - \lambda_{a \s} \right) 
f^{\dagger}_{a \s} (\bk) f_{a \s}(\bk), 
\label{eq:Hf}
\end{equation}
where $\mu$ is the chemical potential, $\e_{a}$ the crystal field and 
$\varepsilon_{a \s}$ is the original fermionic dispersion.  In this single-site 
approximation the renormalization of the dispersion is given by the 
quasiparticle weight $$Z_{a \s} =  \langle O_{a \s} \rangle^2,$$ 
self-consistently determined from the solution of the pseudospin Hamiltonian. 
\begin{eqnarray}
H_{\text{PS}} &=& \sum_{a \, \s}   h_{a \s } O_{a \s}  + \sum \lambda_{a \s}\left( S^z_{a \s} +\frac{1}{2} \right) \nonumber \\
&+& \frac{U^\prime}{2} \left( \sum_{a \s} S^z_{a \s} \right)^2 + J_H \sum_a \left( \sum_{\s} S^z_{a \s} \right)^2  \nonumber \\
&-& \frac{J_H}{2} \sum_\s \left( \sum_{a} S^z_{a \s} \right)^2
\nonumber
\label{eq:HPS}  
\end{eqnarray} 
where $h_{a \s} = \langle O_{a \s}\rangle \sum_\bk \varepsilon_{a \s}(\bk) \langle f^\dagger_{a \s}(\bk)f_{a \s}(\bk)\rangle$.

In the case of a spin and orbital degenerate system without spontaneous breaking 
of the symmetry, as the one discussed in the main text,  $\lambda_{a,\sigma}$, 
$h_{a,\sigma}$ and $Z_{a,\sigma}$ become orbital and spin independent and the 
corresponding indices can be dropped. A convenient choice for $\gamma_{i,a 
\sigma}=\gamma$ in this case is\cite{demediciprb2010} $\gamma= 
\frac{1}{\sqrt{n(1-n)}}-1$ 

In Fig. \ref{fig:supp5} the quasiparticle weight $Z$ calculated within this 
Ising Z$_2$ slave--spin approach for a system 3--degenerate orbitals with $n$ 
electrons per site is compared with the DMFT results from 
\cite{demediciprl2011}, which include the full rotationally invariant Hund 
interaction. Here we use a semicircular density of states, different 
to the square lattice with hopping to first nearest neighbors used in the main text. An overall agreement between the two methods is observed.  The 
Ising slave--spin approach captures the different behaviors of $Z$ observed for 
the whole range of $J_H$ values. Quantitatively, the agreement is quite good for 
$n=1,3$ being, $Z$ and the critical $U_c$ for the Mott transition just slightly 
overestimated, as it is common in slave variable approaches and in the 
Gutzwiller approximation.  At intermediate filling ($n=2$) and large $J_H$, the 
suppression of $Z$ in slave--spin approach is weaker than in DMFT. Moreover $U_c$ is 
underestimated probably due to a more prominent role played in these region of parameters by the pair-hopping and spin-flip 
terms, neglected in the calculation. The qualitative behavior, an 
extended metallic region with reduced coherence, is in any case well captured.

\bibliography{hundmetal}

\begin{thebibliography}{39}
\bibitem[{not({\natexlab{a}})}]{notainteractions}
\bibinfo{note}{Only density-density interactions are included to calculate the
  interaction energy cost.}
\bibitem[{not({\natexlab{b}})}]{notadecoupling}
\bibinfo{note}{A clear evidence of the decoupling produced by the effective
  reduction of the interaction appears for 6 electrons in 5 orbitals. With
  increasing $U$ (at fixed $J_H/U$), the ratio $<\delta n_a \delta
  n_b>/<|\delta n_a|^2>$ shows a minimum at the entrance in the Hund metal.
  That is, while $U'$ and $U'-J_H$ increase, the correlation between electrons
  in different orbitals decreases. This minimum can only be explained by the
  larger (smaller) number of electron pairs interacting with $U'-J_H$ (U')
  induced by the spin polarization. In the 2 electron in 3 orbitals case, to
  our degree of accuracy, the minimum is only observed at large $J_H/U$.}
\end{thebibliography}

\end{document}